 \documentclass[final,1p,times]{elsarticle}

\newcommand{\Msun}{\mbox{$M_{\odot}$}}
\newcommand{\Rsun}{\mbox{$R_{\odot}$}}

\newcommand{\kms}{\mbox{km s$^{-1}$}}
 \usepackage{graphics}
 \usepackage{graphicx}
 \usepackage{epsfig}

\usepackage{amssymb}
\usepackage{amsthm}

\usepackage{lineno}

\journal{New Astronomy}
\begin{document}
\begin{frontmatter}

  \title{A new low-mass eclipsing binary: NSVS\,02502726\tnoteref{t1}}
  \tnotetext[t1]{Based on photometric and spectroscopic observations collected at T\"{U}B\.{I}TAK National Observatory (Turkey)}
  \author[rvt]{\"{O}. ̃\c{C}ak{\i}rl{\i}\corref{cor1}}
  \ead{omur.cakirli@ege.edu.tr}
      \author[rvt]{C. \.{I}bano\v{g}lu}
      \author[rvt]{C. G\"{u}ng\"{o}r}
       
          \cortext[cor1]{Corresponding author}
\address[rvt]{Ege University, Science Faculty, Department of Astronomy and Space Sciences, 35100 Bornova, \.{I}zmir, Turkey
}
                                                                      
\begin{abstract}
We present optical spectroscopy and extensive $RI$ differential photometry of the double-lined eclipsing binary NSVS\,02502726 
(2MASS J08441103+5423473). Simultaneous solution of two-band light curves and radial velocities permits determination of 
precise emprical masses and radii for both components of the system. The analysis indicates that the primary and secondary 
components of NSVS\,02502726 are in a circular orbit with 0.56-day orbital period and have stellar masses of M$_1$=0.714$\pm$0.019 \Msun, 
and M$_2$=0.347$\pm$0.012 \Msun. Both of the components have large radii, being R$_1$=0.645$\pm$0.006 \Rsun, and R$_2$=0.501$\pm$0.005 \Rsun. The 
principal parameters of the  mass and radius of the component stars are found with an accuracy of 3\% and 1\%, respectively. The 
secondary component's radius is significantly larger than model predictions for its mass, similar to what is seen in almost all of 
the other well-studied low-mass stars which belong to double-lined eclipsing binaries. Strong H$_{\alpha}$ emission cores and 
considerable distortion at out-of-eclipse light curve in both $R$ and $I$ bandpasses, presumably due to dark spots on both stars, have 
been taken as an evidence of strong  stellar activity. The distance to system was calculated as 173$\pm$8 pc from the $BVRIJHK$ 
magnitudes. The absolute parameters of the components indicate that both components are close to the zero-age main-sequence. Comparison 
with current stellar evolution models gives an age of 126 $\pm$30  Myr, indicating the stars are in the final 
stages of pre-main-sequence contraction.  
\end{abstract}
\end{frontmatter}

\linenumbers


\section{Introduction}
Most of the investigations made to determine the  fundamental properties of low-mass stars using eclipsing binaries 
indicate a strong discrepancy between theory and observations. Radii measurements  of the low-mass stars can be made 
from the eclipsing binaries plus interferometry of single stars. These measurements clearly indicate that the observed 
radii are generally larger than the predictions by the stellar models. On the other hand the observed temperatures are 
lower than those of predicted by the models.   This discrepancy, the observed larger radii and lower temperatures, is 
generally  explained by the high level  of stellar activity.  

Accurate parameters of low-mass stars are difficult to obtain, with the best source of precise data being double-lined 
eclipsing binaries, but those systems are not only scarce but also intrinsically faint. Therefore their detection is 
slightly difficult. As pointed out by Coughlin and Shaw (2007) only three low-mass double-lined eclipsing binary 
systems  were known before 2003:  CM\,Dra (Lacy,1977 and Metcalfe et al. 1996), YY\,Gem (Leung and Schneider 1978, Torres 
and Ribas 2002) and CU\,Cnc (Delfosse et al. 1999, Ribas 2003).  In the following five years the number of low-mass 
systems has tripled by many variability surveys:   BW5\,V038 (Maceroni and Montalban 2004), TrES-Her 0-07621 (Creevey 
et al. 2005), GU\,Boo (Lopez-Morales and Ribas 2005), 2MASS J05162281+2607387 (Bayles and Orosz 2006), NSVS01031772 
(Lopez-Morales et al. 2006), UNSW-TR-2 A and B (Young et al. 2006), and 2MASS J04463285+1901432 A and B in the open 
cluster NGC\,1647 (Hebb et al. 2006). Thereafter, seven new low-mass eclipsing binaries were discovered by  Coughlin 
and Shaw (2007). Very recently Lopez-Morales et al. (2006) discussed a plausible correlation between the magnetic activity 
levels, the metallicities and the radii of low mass stars which depends on the precise radii measurements of 34 low mass 
stars from the eclipsing binary systems. Ribas and colleagues (Ribas et al. 2003, Morales et al. 2008) have revealed 
that the low-mass stars are systematically larger and cooler than the predictions of theoretical calculations. However, there 
is no significant difference between the observed and theoretical luminosities of the low-mass stars.     

Clearly, the sample size of well-studied low-mass binaries needs to be increased. Since the number of well-studied low-mass 
binaries is still relatively small, observations of additional low mass binaries would be extremely useful. Light variability 
of the star known as NSVS\,02502726 (hereafter NSVS\,0250) was revealed by Wozniak et al.( 2004)  from the Northern Sky 
Variability Survey. Later on it has been discovered to be an eclipsing binary by Coughlin \& Shaw (2007).  Their preliminary 
study shows that NSVS\,0250 consists of two dissimilar low-mass stars with a join apperant visual magnitude of 
V$_{rotse}$=13.41 and an orbital period of  $\sim$0.6 days. Since spectroscopic observations are not available, they 
presented the photometric light curves and preliminary models based only on the light curve analysis. 

In this work we present follow-up photometric and spectroscopic observations of NSVS\,0250 which confirm the low-mass 
nature of the component stars. We derive accurate  fundamental parameters for the component stars and compare our results 
with theoretical evolutionary models.

\section{Observations and reductions}
\subsection{Differential photometry}
We report here new photometry of NSVS\,0250 in the Bessell $R$ and $I$ bands. Both the photometric accuracy (a few millimagnitudes) 
and the phase coverage (over 1\,000 observations) are sufficient to guarantee a reliable determination of the light curve 
parameters. The observations were carried out with the 0.40 m telescope between January and February of 2008 at the 
T\"{U}B\.{I}TAK National Observatory (TUG, located on Mt. Bak{\i}rl{\i}tepe, Antalya in south of Turkey).The telescope is 
equipped with an Apogee 1k$\times$1k CCD (binned 2$\times$2) and standard Bessel $R$ and $I$ filters. 

The instrument with attached camera provides a field-of-view of 11$^{\prime}$.3$\times$11$^{\prime}$.3. NSVS\,0250 is a 
relatively faint target, with not many other objects of similar spectral type or brightness nearby. By placing NSVS\,0250 
very close to center of the CCD to get the highest accuracy, we managed to strategically locate the binary on the chip together 
with two other stars of similar apparent magnitudes. We selected GSC 3798-1250 as a comparison star, located  1$^{\prime}$.479 
away from the target. The check star was GSC 3798-1234, at an angular distance of about 3$^{\prime}$.047 from NSVS\,0250, 
2$^{\prime}$.629 from the comparison star. Both stars passed respective tests for intrinsic photometric variability and proved 
to be stable during time span of our observations. The variable, comparison and check stars are shown in Figure 1. 

\begin{figure}
\includegraphics[width=12cm]{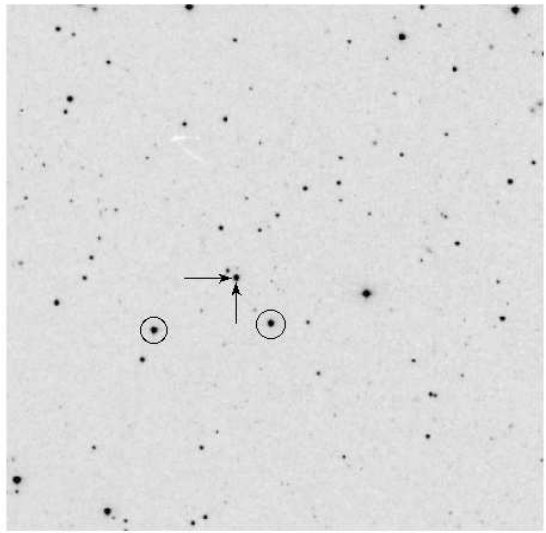}
\caption{A 11$^{\prime}$.3$\times$11$^{\prime}$.3 I band image of the field of NSVS\,0250 showing the two comparison 
stars. The position of NSVS\,0250 is indicated by vectors directed by east and north.
}
\label{Fig. 1.}
\end{figure}

We collected a total of 1190 differential magnitudes in the $R$-band and 1235 in the $I$-band with an exposure time of 10 
seconds. The observations covered the entire orbit of the binary in each filter. Machine-readable copies of the data 
are available in the electronic edition (see supplementary material section of this article). Standard IRAF\footnote{IRAF 
is distributed by the National Optical Observatory, which is operated by the Association of the Universities for Research 
in Astronomy, inc. (AURA) under cooperative agreement with the National Science Foundation} tasks were used to remove the 
electronic bias and to perform the flat-fielding corrections. The IRAF task {\sc imalign} was used to remove the differences 
in the pixel locations of the stellar images and to place all the CCD images on the same relative coordinate systems. The data 
were analyzed using another IRAF task {\sc phot} without taking into account  differential extinction effects due to the 
relatively small angular separation  between the target, comparison and check stars on the sky. 

\subsubsection{Orbital period and ephemeris}
Coughlin \& Shaw (2007) observed seven low-mass detached systems, including NSV0250, with the Southeastern Association for 
Research in Astronomy (SARA) 0.9 m telescope in the Johnson $V$, $R$ and $I$ filters. The first orbital period and zero epoch for 
NSVS\,0250 were determined from these observations. An orbital period of P=0.559772$\pm$0.000007 days, and 
an initial epoch T$_0$(HJD)=2453692.0280$\pm$0.0003 for the mid-primary eclipse were calculated using a least square fit. To 
define the accurate period of NSVS\,0250, We collected times of minima avaliable from the literature and added  8 times of 
mid-eclipse obtained in this study, including 4 primary and 4 secondary. Accurate times for those mid-eclipses were computed 
by applying 6$^{th}$ order polynominal fits to the data during eclipses. The times of mid-eclipse obtained by us are listed 
in Table 1. A linear least squares fit to the obtained so far yields an orbital period of P=0.559755$\pm$0.000001 days, which 
is 1.5 s shorter than that estimated by Coughlin \& Shaw (2007). As the new reference epoch, we have adopted the first time 
of mid-primary eclipse that we observed, i.e. T$_0$(HJD)=2454497.5502$\pm$0.0003. The ephemeris of the system is now,
\begin{equation}
Min\, I\, (HJD)= 24\,54497.5502(3)+0.559755(1)\times E. 
\end{equation}  
Using these light elements we find an average phase difference between mid-primary and mid-secondary eclipses 
$\Delta \phi$=0.4992$\pm$0.0008, which is consistent with a circular orbit.

\begin{table}
\caption{Times of minima measured from the $BVRI$-band light curves. Caution: the first two columns should be shifted to the right.}
\begin{tabular}{cccc}
\hline
E  & Type	&HJD          &O-C \\
\hline
0.50  &II	&54497.2696& -0.0007\\
1.00  &I 	&54497.5502&  0.0000\\
2.50  &II	&54498.3894& -0.0005\\
3.00  &I 	&54498.6701&  0.0003\\
4.00  &I 	&54499.2300&  0.0004\\
4.50  &II	&54499.5085& -0.0010\\
6.50  &II	&54500.6284& -0.0007\\
8.00  &I 	&54501.4690&  0.0002\\
\hline
\end{tabular}
\end{table}

\subsection{Echelle spectroscopy}
Optical spectroscopic observations of NSVS\,0250 (30 spectra) were obtained with the Turkish Faint Object Spectrograph Camera
(TFOSC) instrument attached to the 1.5 m telescope on 3 nights on 21, 22, 23 February 2008 under good seeing conditions. The TFOSC instrument 
equipped with a 2048$^2$ pixel CCD was used. Further details on the telescope and the instrument can be found at 
http://www.tug.tubitak.gov.tr. The wavelength coverege of each spectrum is 4200-8700 \AA~in 11 orders, with a resolving 
power of $\lambda$/$\Delta \lambda$ 6\,000 at 6563 \AA~and an average signal-to-noise ratio (S/N) of $\sim$140. We also 
obtained a high S/N spectrum of the M dwarf GJ 410 (M0 V) and GJ 361 (M1.5 V) for use as a template in derivation of the radial 
velocities (Nidever et al. 2002). By using a real star as template we avoid the problems that the low-mass stellar atmosphere models 
have been reproducing some spectral features of the stars. 

The electronic bias was removed from each image and we used the 'crreject' option for cosmic ray removal. This worked very well, and
the resulting spectra were largely free from cosmic rays. The echelle spectra were extracted and wavelength calibrated by
using FeAr lamp source with help of the IRAF {\sc echelle} package. 
 
The stability of the instrument was checked by cross correlating the spectra of the standard star against each other using the {\sc fxcor}
task in IRAF. The standard deviation of the differences between the velocities measured using fxcor and the velocities in Nidever et al. (2002)
was about 1.1 \kms.

\begin{figure}
\includegraphics[width=11cm]{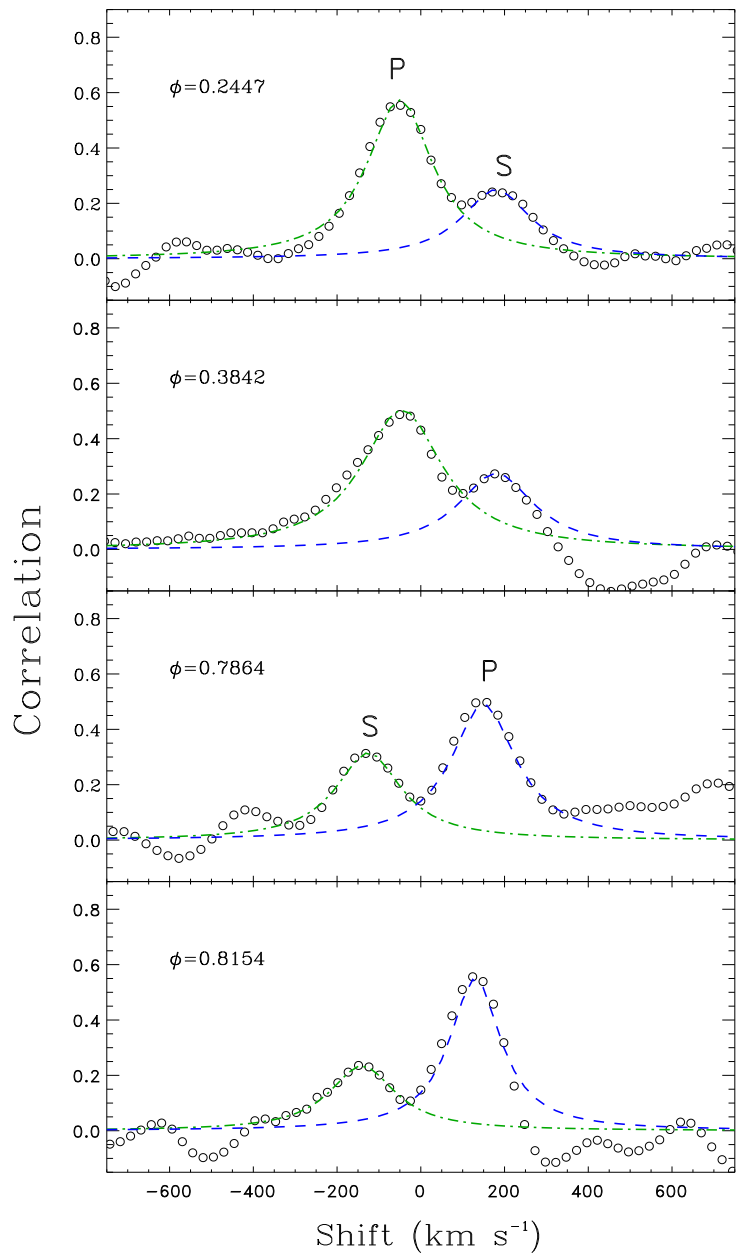}
\caption{Sample of Cross Correlation Functions (CCFs) between NSVS\,0250 and the radial velocity template 
spectrum around the first and second quadrature.}
\label{Fig. CCF}
\end{figure}

\begin{figure}
\includegraphics[width=9cm]{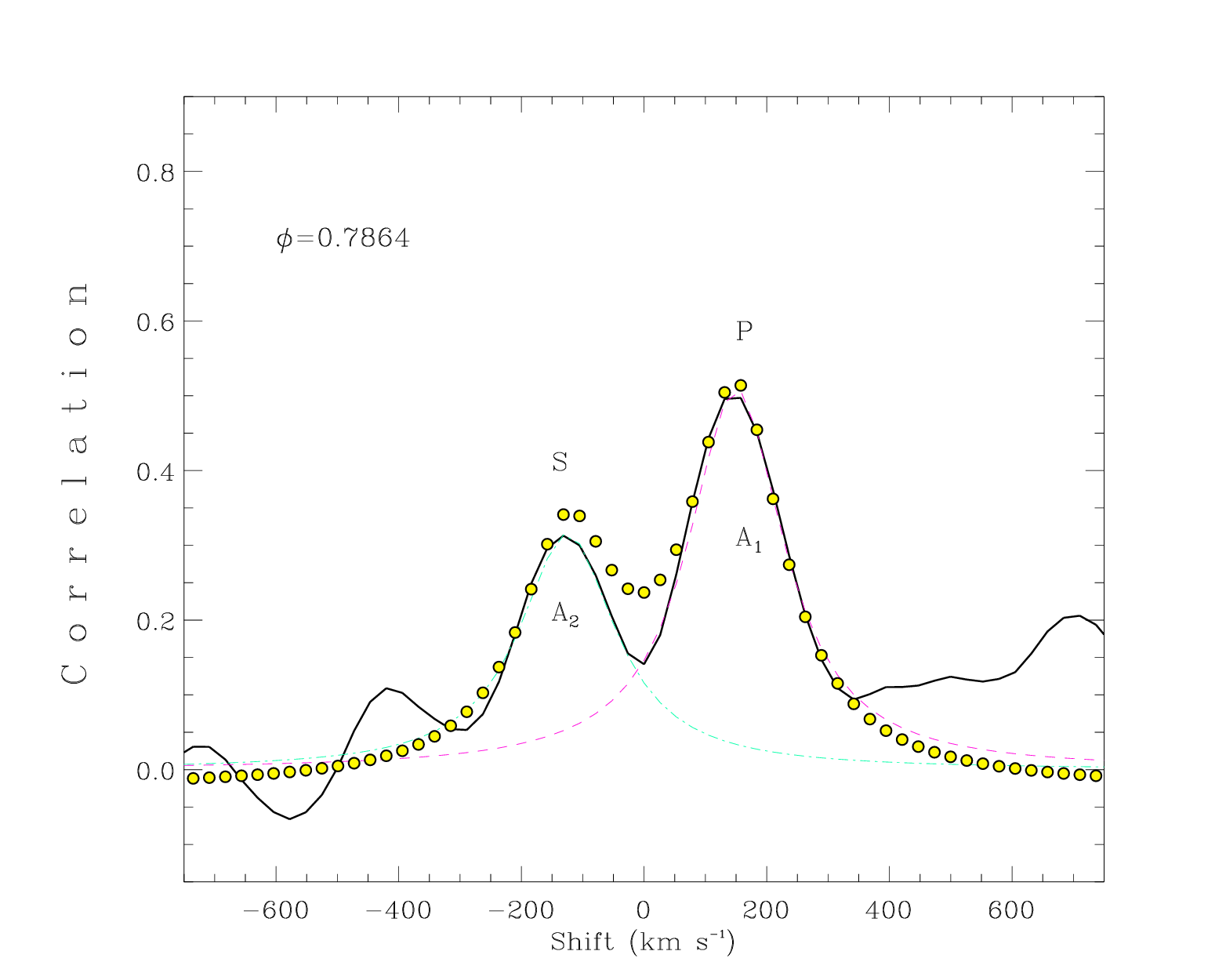}
\caption{Sample of "deblending" a triple-peaked cross-correlation function (CCF) between NSVS\,0250 and
template spectrum GJ\,410 at $\phi$=0.7833 (showing clear double-lined profile in figure). P, and S refer to
primary, secondary components, respectively. The fit to the observed spectrum for the P and S made up by
two Lorentian function is shown by dots. All radial velocities of the primary and secondary have been measured by excluded
of the deblending effect using this way. The deconvolved two gaussian profiles of the primary and 
secondary component are displayed by dashed and dot-dashed line in the figure.
}
\label{Fig. A_1_alanA_2}
\end{figure}

In the present work, the radial velocities of the components of NSVS\,0250 were derived by means of cross-correlation functions (CCFs) 
using the IRAF task {\sc fxcor} (e.g.\ Tonry \& Davis 1979). We used the spectra of the M dwarfs GJ 410 and GJ 361 as trial 
templates instead of the synthetic spectra, the common procedure for massive stars. The reason behind this decision is that the synthetic
spectra computed using stellar atmosphere models reproduce well the spectral features observed in real stars down to 
0.60-0.65 \Msun (T$_{eff}$ $\leqslant$ 4\,000 K), as is for example the case of the H$_2$O molecule. To avoid the incompletness of 
the models we used the spectra of real stars with spectral type similar to the components of the binary. 

Fig.\,2 shows examples of CCFs at various orbital phases. The two peaks, non-blended, correspond to each component 
of NSVS\,0250. The stronger peaks in each CCF correspond to the more luminous component that have a 
larger weight into the observed spectrum. We adopted a two-Gaussian fit algorithm to resolve cross-correlation peaks near 
the first and second quadratures when spectral lines are visible separetely. 

\begin{table}
\centering
\begin{minipage}{85mm}
\caption{Heliocentric radial velocities of NSVS\,0250. The columns give the heliocentric Julian date, the
orbital phase (according to the ephemeris in Eq.~1), the radial velocities 
of the two components with the corresponding standard deviations.
}
\label{Table 2.}
\begin{tabular}{@{}ccccccccc@{}}
\hline
HJD 2400000+ & Phase & \multicolumn{2}{c}{Star 1 }& \multicolumn{2}{c}{Star 2 } &         	& 			\\
             &       & $V_p$                      & $\sigma$                    & $V_s$   	& $\sigma$	\\
\hline
54518.3468  &0.1530 &-67.7  &6.3  &132.5   &11.8     \\
54518.3805  &0.2132 &-78.7  &4.6  &178.1   & 8.8     \\
54518.4122  &0.2699 &-61.3  &8.8  &177.8   & 9.1     \\
54518.4525  &0.3419 &-67.7  &9.7  &170.5   & 7.0     \\
54518.4980  &0.4232 &-30.9  &15.7 &99.5    & 14.7    \\
54518.5804  &0.5704 &31.9   &18.9 &-67.0   & 10.9    \\
54518.6232  &0.6468 &78.5   &11.1 &-125.7  & 6.6     \\
54519.2611  &0.7864 &91.6   &4.6  &-163.2  & 4.3     \\
54519.2684  &0.7995 &92.4   &8.9  &-154.7  & 6.7     \\
54519.2975  &0.8515 &86.0   &7.1  &-136.7  & 7.7     \\
54519.3416  &0.9302 &55.2   &15.7 &-78.4   & 12.4    \\
54519.4336  &0.0946 &-29.5  &19.9 &114.1   & 18.4    \\
54519.4797  &0.1770 &-71.1  &3.5  &161.8   & 6.7     \\
54519.5191  &0.2473 &-70.7  &5.5  &189.7   & 8.9     \\
54519.5476  &0.2983 &-73.0  &7.8  &179.5   & 11.9    \\
54519.5957  &0.3842 &-45.4  &5.8  &129.4   & 8.8     \\
54519.6400  &0.4633 &-15.3  &14.6 &62.0    & 17.6    \\
54520.3968  &0.8154 &88.6   &6.6  &-146.3  & 7.6     \\
54520.4551  &0.9195 &55.7   &11.2 &-84.8   & 11.5    \\
54520.5454  &0.0808 &-27.2  &19.7 &91.0    & 12.3    \\
54520.5694  &0.1237 &-54.4  &14.4 &114.1   & 9.8     \\
54520.6371  &0.2447 &-77.7  &6.6  &185.1   & 10.8    \\
\hline \\
\end{tabular}
\end{minipage}
\end{table}

Near the quadrature phases, absorption lines of the primary and secondary components of the system can be easily 
recognized in the range between  4200-6800 \AA\ . We limited our analysis to the echelle orders in the  spectral 
domains 4200-6800\,\AA, which include several photospheric absorption lines. We have disregarded very broad lines like 
H$\alpha$, H$\beta$ and H$\gamma$ because their broad wings affect the CCF and lead to large errors. A double-lined Gaussian fit was used to 
disentangle the CCF peaks and determine the RVs of each component. Following the method proposed by Penny et al. (2001) we 
first made two-Gaussian fits of the well separeted CCFs using the deblending procedure in the IRAF routine {\sc splot}. The 
average fitted FWHM is 200$\pm$12, and 190$\pm$10 \kms for the primary and secondary components, respectively. In Fig. 3 we 
show a sample of double-Gaussian fit. Indeed, the shapes and velocities corresponding to the peaks of the CCFs are slightly 
changed. By measuring the areas enclosed by the Lorentian profiles of the spectral lines belonging to the primary (A$_1$) and 
secondary (A$_2$) we estimate the light ratio of the primary star to the secondary as 1.613. Using this light ratio we 
find $\frac{L_1}{(L_1 + L_2)}$=0.617.

The heliocentric RVs for the primary (V$_p$) and the secondary (V$_s$) components are listed in Table\,2, along with the dates 
of observation and the corresponding orbital phases computed with the new ephemeris given in \S 2.1.1 The velocities in that 
table have been corrected to the heliocentric reference system by adopting a radial velocity of -13.9 \kms for the template 
star GJ\,410 (Giese 1991). The RVs listed in Table\,2 are the weighted averages of the values obtained from the cross-correlation 
of orders \#4, \#5, \#6 and \#7 of the target spectra with the corresponding order of the standard star spectrum. The weight 
$W_i = 1/\sigma_i^2$ has been given to each measurement. The standard errors of the weighted means have been calculated on 
the basis of the errors ($\sigma_i$) in the RV values for each order according to the usual formula (e.g.\ Topping 
1972). The $\sigma_i$ values are computed by {\sc fxcor} according to the fitted peak height, as described by 
Tonry \& Davis (1979). The observational points and their error bars are displayed in Fig.\,4 as a function of the 
orbital phase. We measure the semi-major axis $a$=2.939$\pm $0.027 $\Rsun$  and semi-amplitudes of the RVs of more massive 
primary and the less massive secondary components to be $K_1=86\pm3$ km s$^{-1}$ and $K_2=177\pm4$ km s$^{-1}$, respectively.

\begin{figure}
\includegraphics[width=9cm]{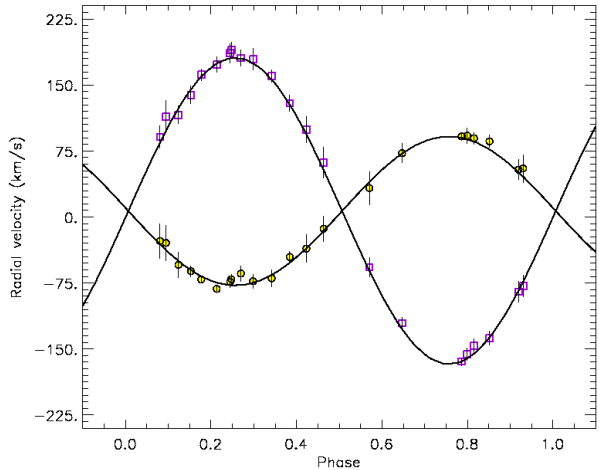}
\caption{Radial velocity curve folded on a period of 0.5598 days, where phase
zero is defined to be at primary mid-eclipse. Symbols with error
bars (error bars are masked by the symbol size in some cases) show the RV
measurements for two components of the system (primary: open circles,
secondary: open squares). 
}
\label{Fig. 4.}
\end{figure}

\section{Analysis}
\subsection{System variability}
The light curves obtained by us in the $I$ and $R$ badpasses are distincly different from the light curves obtained by 
Coughlin \& Shaw (2007) in the same badpasses . Their $VRI$ light curves were shown in Figure 2 (hereafter Fig. 2X) in 
that paper. The light curves are nearly symmetric in shape, showing the system slightly brighter at second quarter than 
at the first quarter. The asymmetry is better distinguished at the shorter wavelengths. The primary and secondary 
eclipses are deeper in the $I$-band, being 0.88 and 0.41 mag, respectively. The $R$ and $I$-band light curves of the system 
obtained by us show considerable out-of-eclipse variability  with an amplitude of about 0.07 mag, presumably due to cool 
spots on one or  both stars. Asymmetries in the light curves are known as  common feature of low-mass eclipsing binaries 
(e.g. the GU Boo light curves shown by Lopez-Morales \& Ribas 2005). However, the activity level in NSVS0250 seems to be 
the same compared to other well-studied low mass binaries (see \S 4.2). If the   light variations at out-of-eclipses are 
due to spots, substantial portion of the surface of one or both components would have been covered with dark spots.

\subsection{Light curve modeling}
We modeled the light curves and radial velocities using the Wilson-Devinney (WD) code implemented into the PHOEBE package 
by Prsa \& Zwitter (2005). The modified version of this program allows one to calculate radial velocities and multiple light 
curves of stars simultaneously, either for a circular or an eccentric orbit, in a graphical environment. In this model the code was 
set in Mode-2 for detached binaries with no constraints on their surface potentials. The simplest assumptions were used for 
modeling the stellar parameters: the stars were considered to be black bodies and the approximate reflection model (MREF=1) 
was adopted. The gravity darkening exponents were set 0.32 according to the mean stellar temperatures given by 
Claret (2000). The bolomeric albedos of 0.5 for each star were taken, which correspond to stars with convective envelopes. 
We assume a circular orbit with synchronous rotation for both stars.   

Since the  light curve is asymmetric in shape we attributed it to spots (either bright and dark) on one or both components. The 
spots in the {\sc PHOEBE} are parameterized in the same way as in the Wilson-Devinney (1971) code. They are circular regions 
specified by four parameters: the "temperature factor" T$_f$, the "latitude" of the spot center, the "longitude" of the spot 
center, and the angular radius of the spot. Bright spots have T$_f$ $>$ 1 and dark spots have T$_f$ $<$ 1.

\begin{figure}
\includegraphics[width=9cm]{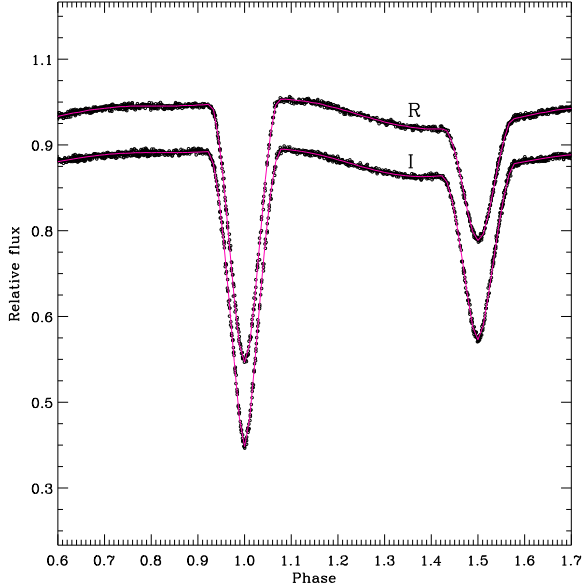}
\caption{The light curve model (solid lines) composed with our I and R-band photometry.
}
\label{Fig. 5}
\end{figure}

The effective temperature of the primary component of the eclipsing pair was derived from the calibrations of 
Drilling \& Landolt (2000) using preliminary estimate of its mass. The primary component of the eclipsing pair seems to 
have a mass of 0.71 M$_{\odot}$  which corresponds to a spectral type of about K3 with an effective temperature of 
4650 K. On the other hand, the infrared colours of NSVS\,0250 given in the 2MASS catalogue (Skrutskie 1999) as  
$H-K$=0.154 and $J-H$=0.584 mag which correspond to an effective temperature of 4300 K in the infrared colors-effective 
temperature relations given by Tokunaga (2000). Therefore we started light curve analysis with an effective temperature 
of the hotter component of 4\,300 K.

We followed two steps to determine the binary parameters. First, the inclination, the effective temperature of the secondary 
star (T$_2$), and potentials (stellar radii) were  iteratively adjusted within {\sc PHOEBE} until both the depths and duration 
of the eclipses are matched with the observed $R$ and $I$ light curves. The iteration was carried out automatically
until convergence is obtained,i.e.  as the set of parameters for which the differential corrections were 
smaller than the probable errors. 

Since the light curve of the system has a wave-like distortion we tried to represent it by cool spots on the components. The shape 
of the distortion curve depends on the number of spots, locations and sizes on the stars' surface. Hundreds of trials indicated that  one 
spot on the primary and two spots on the secondary component could reproduce the distortion on the light curve. The temperature of the 
secondary component, T$_2$=3\,620$\pm$205, has been derived from the temperature ratio provided by the analysis of the light curve. The 
effective temperature of 3\,620 K for the less massive component corresponds to an M2 star.

We fit our photometric data with model light curves using the spot parameters given in Table 3. No satisfactory fit was possible without 
starspots, moderate cool spots were added one at a time, with typical temperature factors, T$_f$=$\frac{T_{spot}}{T_{photophere}}$, of about 
0.85, 0.78 and 0.94.  In Fig. 5, the computed light curves with one spot on the primary and two spots on the secondary star,  given in 
Table 3,  are compared with observed $R-$ and $I$-band light curves.

\begin{table}
\centering
\caption{Properties of the NSVS\,0250 components.}
\begin{tabular}{llr}
\hline
\#&Parameters & Value  \\
\hline
&P$^a$ (days)							&0.559755								\\
&T$_0^a$ (HJD) {\footnotesize (Min I)}	&24\,54497.5502							\\
&$\gamma$ (km s$^{-1}$)	       			&3.15$\pm$0.44							\\
&q=$\frac{m_2}{m_1}$					&0.486$\pm$0.010						\\
&$i^{o}$			               	&87$\pm$1								\\
&$a$ (R$_{\odot}$)						&2.914$\pm$0.026						\\
&K$_1$ (km s$^{-1}$)					&86$\pm$3								\\
&K$_2$ (km s$^{-1}$)					&177$\pm$4								\\
&${L_{1}}/{(L_{1}+L_{2})_{R}}$ 			&0.6200$\pm$0.0010						\\
&${L_{1}}/{(L_{1}+L_{2})_{I}}$ 			&0.3800$\pm$0.0009						\\
&$\Omega_1$								&4.8037$\pm$0.0021						\\
&$\Omega_2$								&3.8624$\pm$0.0034						\\				
&r$_1$									&0.2312$\pm$0.0005						\\
&r$_2$									&0.2619$\pm$0.0008						\\
&T$_{eff_1}$ (K)						&4\,300[Fix]							\\
&T$_{eff_2}$ (K)						&3\,620$\pm$205							\\
&$rms$									&0.044									\\				
Spot1									&$Star~location$ 				&Primary	\\
parameters								&Latitude (deg)					&37			\\
										&Longitude (deg)				&254		\\
										&Angular radius (deg)			&38			\\
										&T$_{spot}$/T$_{photosphere}$	&0.85		\\				             		     
Spot2									&$Star~location$				&Secondary	\\
parameters								& Latitude (deg)				&39			\\
										&Longitude (deg)				&261		\\
										&Angular radius (deg)			&54			\\
										&T$_{spot}$/T$_{photosphere}$	&0.78		\\
Spot3									&$Star~location$				&Secondary	\\
parameters								& Latitude (deg)				&22			\\
										&Longitude (deg)				&141		\\
										&Angular radius (deg)			&57			\\
										&T$_{spot}$/T$_{photosphere}$	&0.94		\\
\hline
\end{tabular}
\begin{list}{}{}
\item[$^{a}$]{\small See \S 2.1.1}
\end{list}
\end{table}

\section{Summary and conclusion}
\subsection{Absolute parameters of the components}
Combining the parameters of the photometric and spectroscopic orbital solutions we derived absolute parameters of the 
stars. The standard deviations of the parameters have been determined by JKTABSDIM\footnote{This can be obtained from http://http://www.astro.keele.ac.uk/$\sim$jkt/codes.html} code, which calculates distance and other physical parameters 
using several different sources of bolometric corrections (Soutworth et al. 2005a). The best fitting parameters are listed 
in Table 3 together with their standard deviations. 

\begin{table}
\centering
\caption{The absolute parameters of the components of NSVS0250.}
\begin{tabular}{lcc}
\hline
Parameter&Primary & Secondary  \\
\hline
Mass (M$_{\odot}$)					&0.714$\pm$0.019 	& 0.347$\pm$0.012		\\
Radius (R$_{\odot}$)				&0.674$\pm$0.006 	& 0.763$\pm$0.007		\\
Temperature (K)						&4\,300$\pm$200 	& 3\,620$\pm$205		\\
Luminosity (L$_{\odot}$)			&0.139$\pm$0.014 	& 0.090$\pm$0.010		\\
$log$\,$g$($cgs$)					&4.635$\pm$0.004 	& 4.213$\pm$0.008		\\
M$_{bol}$ ($mag.$)					&6.88$\pm$0.14		& 7.35$\pm$0.12 		\\
M$_V$ ($mag.$)						&7.73$\pm$0.13		& 9.11$\pm$0.14 		\\
Distance (pc)                 & \multicolumn{2}{c}{$173\pm8$}              	\\
\hline
\end{tabular}
\end{table}

The ratio of the temperatures of the stars is consistent with their mass and radius ratios. However, a more accurate determination 
of the mean absolute temperature of NSVS\,0250's secondary is still necessary. Note that the potentially large uncertainity in the 
adopted effective temperature of the primary and calculated for the secondary has no impact on the accuracy of the determined 
absolute dimensions. For example, tha radii of the stars, which are obtained from the light curve modelling, suffer in appreciable 
changes when T$_{eff}$ values moderate uncertainity about the mean one adopted.

The luminosity and absolute bolometric magnitudes M$_{bol}$ of the stars in Table 4 were computed from their effective temperatures 
and their radii. Since low-mass stars radiates more energy at the loger wavelengths we used $RIJHK$ magnitudes given by 
Coughlin \& Shaw (2007). Applying $UBVRIJHKL$ brightness-T$_{eff}$ relations given by Kervella et al. (2004) we calculated the 
distance to NSVS\,0250 as $d$=173$\pm$8 pc. 

The mean light contribution of the primary star $\frac{L_1}{(L_1 + L_2)}$=0.62 obtained from the $R$-band light curve analysis is in 
agreement with that estimated from the FWHM as 0.62. On the other hand we find a light contribution of the primary component as 0.61 
using the absolute parameters given in Table 4. This indicates that light contributions of the spotted primary component, computed by 
three different ways, are very close to each other.

\begin{figure}
\includegraphics[width=9cm]{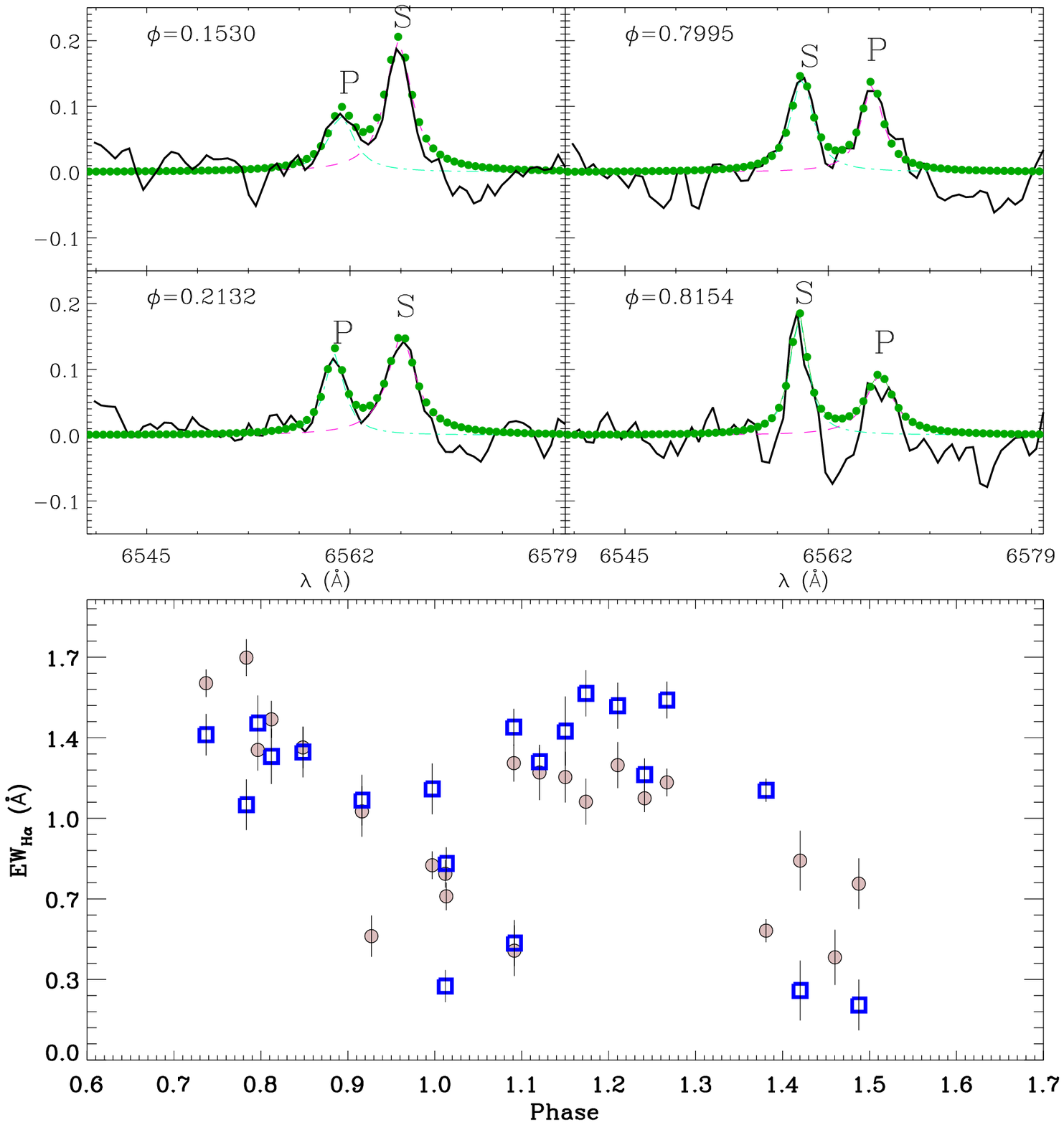}
\caption{Sample deblending of the double-peaked H$_{\alpha}$ emission spectra of NSVS\,0250 (top panels). The fit to the 
observed spectrum (continuous line) made up by two Lorentian function is shown by dots. The deconvolved emission profiles 
of the primary and secondary component are displayed by dashed and dot-dashed line, respectively. In the bottom panel we 
show equivalent variation of the H$_{\alpha}$-line versus the orbital phase. The symbols square and circle denote the EWs 
of the secondary and primary component, respectively. 
}
\label{Fig. 6.}
\end{figure}

\subsection{H$_{\alpha}$ emission profiles}
The H$_{\alpha}$ line is an important indicator of photospheric and chromospheric activity in the low-mass stars. Very
active binaries show H$_{\alpha}$ emission above the continuum (e.g. SDSS-MEB-1, Blake et al. 2007); in less active stars 
a filled-in absorption line is observed. For some objects the H$_{\alpha}$ line goes from filled-in absorption to emission 
during flare events as shown in YY Gem (Young et al. 1989). In the spectra of NSVS\,0250 collected by us clearly indicated 
strong emission in H$_{\alpha}$ for both components above the continuum in some orbital phases. However, in some orbital 
phases the  H$_{\alpha}$ lines appear to be very shallow absorption, i.e. filled-in absorption line, below the 
continuum. In Fig. 6 we show H$_{\alpha}$ emission features in some orbital phases together with their equivalent width 
variation (bottom panel) as a function orbital phase. We find a clear evidence of the H$_{\alpha}$-line equivalent width 
(EW) variation with the orbital phase, indicating direct emission with larger EW when the spotted areas are visible. Such 
correlation between the EWs of H$_{\alpha}$-line and orbital phase is known for chromospherically active RS CVn-type stars 
for a long time.  

Our observations of NSVS\,0250 are well distributed in orbital phase and were obtained at a spectral resolution 0.9 \AA. Both 
components display H$_{\alpha}$ emission cores and the more massive component usually shows weaker emission. Four spectra of 
NSVS\,0250 in the deblending of a double-peaked H$_{\alpha}$ region show emission core from both stars in any case for the
comparable intensities of the components. The radial velocities of the components could be measured from the deblending 
H$_{\alpha}$ emission lines, near the orbital phases of 0.25 and 0.75. These RVs are very close to those obtained from the 
absorption lines, indicating that the H$_{\alpha}$ emission lines are formed in a region close to the stars' surface.

The relevant result of the simultaneous photometric and H$_{\alpha}$ monitoring is that the less massive and cooler star 
appears also as the more active at a chromospheric level, since it has a larger H$_{\alpha}$-line's EWs at this 
epoch. Therefore, one can conclude that  the secondary component should be  more heavily spotted, which is confirmed by the light curve analysis. 

\begin{figure}
\includegraphics[width=9cm]{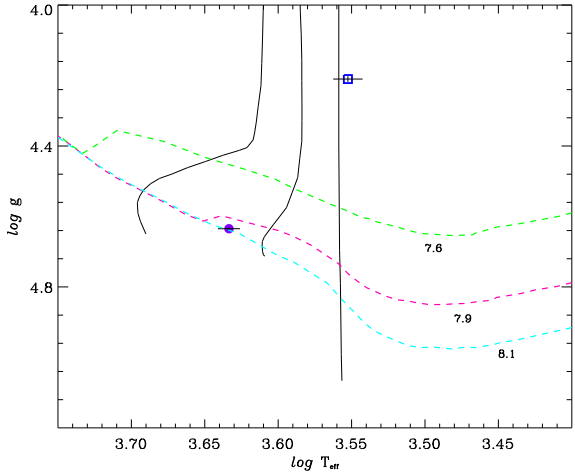}
\caption{Location of the primary (filled circle) and secondary (open square) stars in the theoretical H-R diagram. The 
components of NSVS\,0250 compared to mass-tracks and isochrones from Palla and Stahler (1999) models. The solid lines are 
mass-tracks of masses 0.4, 0.6, 0.8 \Msun. The dashed lines are [M/H]=0 isochrones of ages $log~t$=7.6, 7.9, and 8.1 
years. }
\label{Fig. 7.}
\end{figure}

\subsection{HR diagram}
We can now compare the fundamental stellar parameters derived from the orbital solution (Table 4) inferred currently available theoretical 
evolutionary models. This comparison may easily be done by comparing location of a star in the log $T_{eff}$ vs. log $g$ or mass-radius 
diagrams formed by theoretical predictions. As an example, Fig. 7 shows the locations of the primary and secondary stars  on the theoretical 
HR diagram, i.e. log $T_{eff}$ vs. log $g$ . The evolutionary tracks for the pre-main sequence stars with masses of 0.8, 0.6 and 0.4 M$_{\odot}$ 
taken from Palla and Stahler (1999) are also plotted for comparison. The location of the primary component in the log $T_{eff}$ vs. log $g$ diagram 
is in agreement with theoretical prediction of a 0.7 M$_{\odot}$. Using the solar metallicity models, we find a best-fit age for the primary star 
of $\sim 126\pm$30  Myr. The uncertainity of the age has been estimated from the errors in the effective temperature and surface gravity of 
the primary star. On contrary the secondary component seems to have an age of about 10 Myr, indicating too young with respect to the primary 
star.  This contradiction may be arisen from the very large radius, therefore smaller surface gravity, of the secondary component.   On the other 
hand, when compared the locations of the components in the log $T_{eff}$ vs. log $L$ diagram, while the primary component appears to be on 
the evolutionary track of a  0.7 M$_{\odot}$ the  secondary star is seen to be having higher luminosity, therefore younger.  Such tests of low-mass 
stellar models have been carried out by a number of authors in the past. Torres \& Ribas (2002) and Ribas (2003) have systematically pointed out 
a discrepancy between the stellar radii predicted by theory and the observations. Model calculations appear to underestimate 
stellar radii by $\sim$ 10 \%, which is a highly significant difference given the observational uncertainities. Recently, Ribas 
et al. (2003) and Morales et al. (2008) made a comparison between the mass-radius and mass-$T_{eff}$ relations predicted by 
models and the observational data for stars below 1 M$_{\odot}$ from detached eclipsing binaries. They conclude that current 
stellar models predict radii for low-mass stars 10\% smaller than measured. Furthermore the computed effective temperatures 
are 5\% larger, while the luminosities are in agreement. While the radius of the more massive component is in agreement with the 
stars with same masses, the secondary component seems to have 1.5 times larger with respect to its mass than predicted by the stellar 
theory for an age of the primary star (see Morales 2007, Figure 1). This is the reason why the location of the secondary component 
does deviate from the predicted in the log $T_{eff}$ vs. log $g$ diagram. Morales et al. (2008) indicate that chromospherically 
active stars are cooler and larger than the inactive stars of similar luminosity. Even if the use of log $T_{eff}$ vs. log $L$ 
diagram, the cooler temperatures will of course lead to infer younger ages even the log $T_{eff}$ vs. log $L$ diagram is used. 

The heliocentric space velocity components of NSVS0250 were computed from its position, radial velocity ($\gamma$), distance 
($d$), and proper motion. The latter were retrived from the $2MASS$ catalogue ($\mu_{\alpha}$, $\mu_{\delta}$). The resulting 
space velocity components are (U,V,W)\footnote{According to our convention, positive values of U, V, and W indicate velocities 
towards the galactic center, galactic rotation and north galactic pole, respectively.}; U=-1.7$\pm$0.2 \kms, V=1.6$\pm$0.3 \kms, 
W=2.6$\pm$0.1 \kms, which correspond to a total space velocity of S=4.1$\pm$0.9 \kms. We can infer from those 
space velocities that NSVS\,0250 should not be an older star. Therefore, we conclude that the system seems to have an age of about 
126 Myr.  Both components are in the final stages of the pre-main-sequence contraction.   
            
\section{Acknowledgements}
The authors acknowledge generous allotments of observing time at TUBITAK National Observatory of Turkey. We also 
wish to thank the Turkish Scientific and Technical Research Council for supporting this work through grant Nr. 108T210 
and  EB{\.I}LTEM Ege University Science Foundation Project No:08/B\.{I}L/0.27. This research has been made use of the 
ADS--CDS databases, operated at the CDS, Strasbourg, France.

\end{document}